\documentclass[prd,twocolumn,nofootinbib]{revtex4}
\usepackage{graphicx, epsfig}
\usepackage{color}
\usepackage{mathrsfs}
\usepackage{bm}
\usepackage{amsmath,amssymb}% for \eqref
\usepackage{mathrsfs}
\usepackage[caption=false]{subfig}
\usepackage{hyperref}
\usepackage[normalem]{ulem}
\usepackage{mathtools}
\usepackage[x11names]{xcolor}

\definecolor{fashionfuchsia}{rgb}{0.96, 0.0, 0.63}
\colorlet{no_so_fashion_purple}{blue!50!red}

\newcommand{\be}{\begin{equation}}
\newcommand{\ee}{\end{equation}}
\newcommand{\ba}{\begin{eqnarray}}
\newcommand{\ea}{\end{eqnarray}}

\newcommand{\nn}{\nonumber}

\newcommand{\hatn}{\hat n}
\newcommand{\hatk}{\hat k}
\newcommand{\hatxi}{\hat \xi}
\newcommand{\hatz}{\hat z}

\newcommand{\bfk}{{\bf k}}
\newcommand{\bfp}{{\bf p}}

\newcommand{\bfx}{{\bf x}}

\begin{document}
\title{Stability analysis of non-Abelian electric fields}
\author{Jude Pereira\footnote{jperei10@asu.edu}, Tanmay Vachaspati\footnote{tvachasp@asu.edu}}
\affiliation{
Physics Department, Arizona State University, Tempe,  Arizona 85287, USA.
}

\begin{abstract}
We study the stability of fluctuations around a homogeneous non-Abelian electric
field background that is of a form that is protected from Schwinger pair production. 
Our analysis identifies the unstable modes and we find a limiting 
set of parameters for which there are no instabilities. We discuss potential implications 
of our analysis for confining strings in non-Abelian gauge theories.
\end{abstract}

\maketitle

\section{Introduction}
\label{introduction}

In pure non-Abelian gauge theory, the gauge fields carry non-Abelian electric charge.
Hence a non-Abelian electric field is susceptible to decay via the Schwinger pair production
of gauge field quanta 
\cite{Matinyan:1976mp,Brown:1979bv,Yildiz:1979vv,Ambjorn:1981qc,Ambjorn:1982nd,
Cooper:2005rk,Nayak:2005yv,
Cooper:2008vy,Nair:2010ea,Ragsdale:2017wgi,Cardona:2021ovn}. 
However, confining electric field tube configurations
do not decay, leading to a quandary -- how are electric flux tubes stable to
Schwinger pair production?
This question was raised and investigated in~\cite{Vachaspati:2022ktr}. The essential idea 
is that there are many gauge inequivalent ways of constructing an electric field in non-Abelian 
theory~\cite{Brown:1979bv}.
The straight-forward embedding of the Maxwell gauge field in the non-Abelian theory is
indeed unstable to Schwinger pair production. However, other inequivalent gauge fields, 
that nonetheless produce the same electric field, are protected against quantum dissipation. 
Such gauge field configurations are candidates for describing confining electric flux tubes.

In a quantum theory, there will be fluctuations about the electric field background and these
fluctuations will ultimately be quantized. It is therefore of interest to determine the fluctuation
eigenfrequencies and eigenmodes, and especially to determine if there are any unstable
fluctuations. The question has been addressed in Refs.~\cite{Bazak:2021xay,Bazak:2022khg} 
for a homogeneous electric field where a number of unstable modes were 
found\footnote{Stability analyses of non-Abelian fields are also of interest in
heavy ion collisions (see for example Ref.~\cite{Berges:2020fwq}.)}.
However those analyses did not eliminate fluctuations that were inconsistent with 
their adopted gauge conditions; neither did they account for extra conditions imposed 
by the reality of the gauge fields. 
Indeed we shall see that these conditions are critical for the stability analysis.
The earlier analyses were also limited to either a special point in the parameter space of the
background electric field~\cite{Bazak:2021xay} or to only the zero momentum 
modes~\cite{Bazak:2022khg}.

We start by describing the homogeneous electric field in SU(2) non-Abelian gauge theory
in Sec.~\ref{Ebackground}. Then we consider small fluctuations around the background 
electric field in Sec.~\ref{perturbations}, expand the fluctuations in modes in 
Sec.~\ref{modeexpansion}. The modes get classified according to whether they
are longitudinal or transverse, and whether they are orthogonal to the electric
field. The transverse-orthogonal (TO) modes are discussed in Sec.~\ref{beta} 
while the transverse-nonorthogonal (TN) and longitudinal (L) modes are
discussed together in Sec.~\ref{alphagamma} as they are coupled. Unstable TO 
modes are found to exist in the infrared and
depend on the parameters entering the background configuration, and an interesting
limit is found for which the unstable TO modes are absent.
The analysis for the TN and L modes is significantly more complicated and we limit ourselves 
to some special cases, for example large or small wavenumbers, and for wave vectors parallel
and orthogonal to the electric field. Our results again show some unstable modes
in the infrared and once more, just as in the case of TO modes, we find that there are no 
unstable modes in the special limit of background parameters.

Our results are summarized in Sec.~\ref{conclusions} where we also discuss other ideas.
The reader not interested in the technicalities of the analysis can find details about
the electric field background in Sec.~\ref{Ebackground} and then proceed to the 
conclusions in Sec.~\ref{conclusions}.

\section{Electric field background}
\label{Ebackground}

Consider the $SU(2)$ pure gauge theory,
\be
L = - \frac{1}{4g^2} W_{\mu\nu}^a W^{\mu\nu a} + j_\mu^a W^{\mu a}
\label{Lag}
\ee
where $g$ is the gauge coupling, $\mu, \nu = 0,1,2,3$ are Lorentz indices and $a=1,2,3$ 
is the color index. The current $j_\mu^a$ is an external current which will be specified below.
The field strength $W_{\mu\nu}^a$ is given in terms of the gauge potential $W_\mu^a$ by,
\be
W_{\mu\nu}^a \equiv \partial_\mu W_\nu^a -\partial_\mu W_\nu^a + \epsilon^{abc} W_\mu^b W_\nu^c .
\label{Wmunu}
\ee 
The gauge field equations of motion are
\be
{\cal D}_\nu W^{\mu \nu a} = j^{\mu a}
\label{eom}
\ee
where
\be
{\cal D}_\nu W^{\mu \nu a} \equiv \partial_\nu W^{\mu \nu a} + \epsilon^{abc}W_\nu^b W^{\mu \nu c} .
\label{covderiv}
\ee

We wish to consider the stability of a class of gauge fields that give rise to homogeneous
electric fields that we treat as a background. The gauge fields are 
\be
A_\mu^\pm \equiv A_\mu^1 \pm i A_\mu^2 = -\epsilon e^{\pm i\Omega t} \partial_\mu z, \ \ A_\mu^3 =0
\label{Amu}
\ee
where $\epsilon$ and $\Omega$ are parameters that label members of the class, and
$z$ is the spatial $z$ coordinate. The electric field is gauge equivalent to~\cite{Vachaspati:2022ktr},
\be
E^a_i = \epsilon \Omega \delta^{a3} \delta_{iz}
\label{Eai}
\ee
and the amplitude of the electric field is
\be
E = \epsilon \Omega
\label{Eamp}
\ee
As shown in Ref.~\cite{Brown:1979bv}, gauge fields with distinct values of $\Omega^2$, even 
for the same value of $E$, are gauge inequivalent.

In the two dimensional parameter space $(\epsilon,\Omega)$, the electric field is constant
whenever $\epsilon\Omega$ is constant. We will find that the limit $\epsilon \to 0$,
$\Omega \to \infty$ but with $E=\epsilon\Omega$ held constant to be of interest from
the point of view of stability.

The external currents $j_\mu^a$ in \eqref{Lag} are chosen such that the background is a 
solution of the classical equations of motion. Therefore
\be
j^{\mu a} = {\cal D}^{(A)}_\nu A^{\mu \nu a}
\ee
which gives
\be
j_\mu^\pm \equiv j_\mu^1 \pm i j_\mu^2 
= - \frac{\epsilon \Omega^2}{2} e^{\pm i\Omega t} \partial_\mu z, \ \ j_\mu^3=0.
\ee
For the purposes of the stability analysis we simply assume that this is an external current,
though it is possible that the currents can arise semiclassically as discussed in 
Ref.~\cite{Vachaspati:2022ktr} and summarized in Sec.~\ref{conclusions}.

\section{Fluctuations}
\label{perturbations}

We now consider small perturbations around the background,
\be
W_\mu^a = A_\mu^a + q_\mu^a.
\label{WAq}
\ee
Inserting this into the equations of motion, \eqref{eom}, and working to linear order
in the perturbations $q_\mu^a$ we get,
\be
\partial_\nu q^{\mu\nu a} + \epsilon^{abc} ( A_\nu^b q^{\mu\nu c} + q_\nu^b A^{\mu\nu c} ) = 0
\label{qeq}
\ee
where
\be
q_{\mu\nu}^a = \partial_\mu q_\nu^a - \partial_\nu q_\mu^a 
      + \epsilon^{abc} (A_\mu^b q_\nu^c - A_\nu^b q_\mu^c )
\ee
and 
\be
A_{\mu\nu}^a = \partial_\mu A_\nu^a - \partial_\nu A_\mu^a +\epsilon^{abc} A_\mu^b A_\nu^c
\ee
which, with our chosen background,
\be
A^\pm_\mu = -\epsilon  e^{\pm i \Omega t} \, \partial_\mu z , \ \ 
A^3_\mu =0,
\label{bkgndpm}
\ee
gives
\be
A_{\mu\nu}^\pm = \pm i E e^{\pm i\Omega t} 
( \partial_\mu z \partial_\nu t  - \partial_\nu z \partial_\mu t ), \ \ A_{\mu\nu}^3=0
\ee
We will be adopting temporal gauge ($W_0^a=0$), so $q_0^a = 0$.

\section{Mode Expansion}
\label{modeexpansion}

We first define fluctuations in a ``rotating frame'', $Q^a_i$, as follows,
\be
q^1_i + iq^2_i \equiv e^{i\Omega t} (Q^1_i + i Q^2_i), \ \ q^3_i \equiv Q^3_i
\ee
where $Q^a_i$ are real. Next we expand $Q_i^a$ in spatial and temporal Fourier 
modes as follows,
\ba
Q_i^a &=&  \int \frac{d^3k}{(2\pi)^3} e^{-i\omega_\bfk t}  e^{i \bfk \cdot \bfx} p_{i,\bfk}^a
\label{FT}
\ea
where $p_{i,\bfk}^a$ are the Fourier amplitudes.

While $\omega_\bfk$ can in general be complex, the reality of the fields $Q_i^a$ 
constrain physical values of $\omega_\bfk$ and $p_i^a$ to satisfy,
\be
\omega_\bfk^* = - \omega_{-\bfk}, \ \  (p^a_{i,\bfk} )^* = p^a_{i,-\bfk}.
\label{reality}
\ee
In what follows we will consider a single $\bfk$ mode and drop the $\bfk$ subscripts,
{\it e.g.} we write $\omega_\bfk$ simply as $\omega$.

Inserting the Fourier expansion into \eqref{qeq} gives 3 constraint equations (the Gauss
constraints) and 9 equations of motion for the 9 components of $p_i^a$. The constraints
are,
\ba
\omega \bfk\cdot\bfp^1-i\Omega\bfk\cdot\bfp^2+\epsilon\Omega p_z^3 &=& 0 
\label{consp1} \\
\omega \bfk\cdot\bfp^2 + i\Omega\bfk\cdot\bfp^1-i\epsilon\omega p_z^3 &=& 0 
\label{consp2}\\
\omega \bfk\cdot\bfp^3 + i\epsilon\omega p_z^2 - 2\epsilon\Omega p_z^1 &=& 0
\label{consp3}
\ea
and the equations of motion are,
\ba
&& \hskip -1.5 cm
(-\omega^2+k^2-\Omega^2) \bfp^1 + i2\omega\Omega \bfp^2 - (\bfk\cdot\bfp^1)\bfk = 0 
\label{eomp1}\\
&&\hskip -1.5 cm
(-\omega^2+k^2-\Omega^2+\epsilon^2) \bfp^2 - i2\omega\Omega \bfp^1 - (\bfk\cdot\bfp^2)\bfk  \nn\\
&& \hskip -0.5 cm
-\epsilon (-i\bfk\cdot\bfp^3+\epsilon p_z^2) {\hat z} + i\epsilon p_z^3 \bfk - i 2\epsilon k_z \bfp^3 = 0 
\label{eomp2}\\
&& \hskip -1 cm
(-\omega^2+k^2+\epsilon^2) \bfp^3 - (\bfk\cdot\bfp^3)\bfk - i \epsilon (\bfk\cdot\bfp^2){\hat z} \nn\\
&& \hskip 1.5 cm
+ i2\epsilon k_z \bfp^2 - i\epsilon p_z^2 \bfk - \epsilon^2 p_z^3 {\hat z}= 0
\label{eomp3}
\ea
where we have now employed the vector notation: $\bfp^a = (p^a_1,p^a_2,p^a_3)$
and $\bfk \cdot \bfp^a = k_i p^a_i$.
It is straightforward to check that any solution of Eqs.~\eqref{consp1}-\eqref{eomp3}
with $\omega_\bfk^*=-\omega_{-\bfk}$ will also satisfy $(p^a_{i,\bfk} )^* = p^a_{i,-\bfk}$,
{\it i.e.} Eqs.~\eqref{consp1}-\eqref{eomp3} are consistent with the reality conditions in 
\eqref{reality}.

\begin{figure}
\includegraphics[width=0.40\textwidth,angle=0]{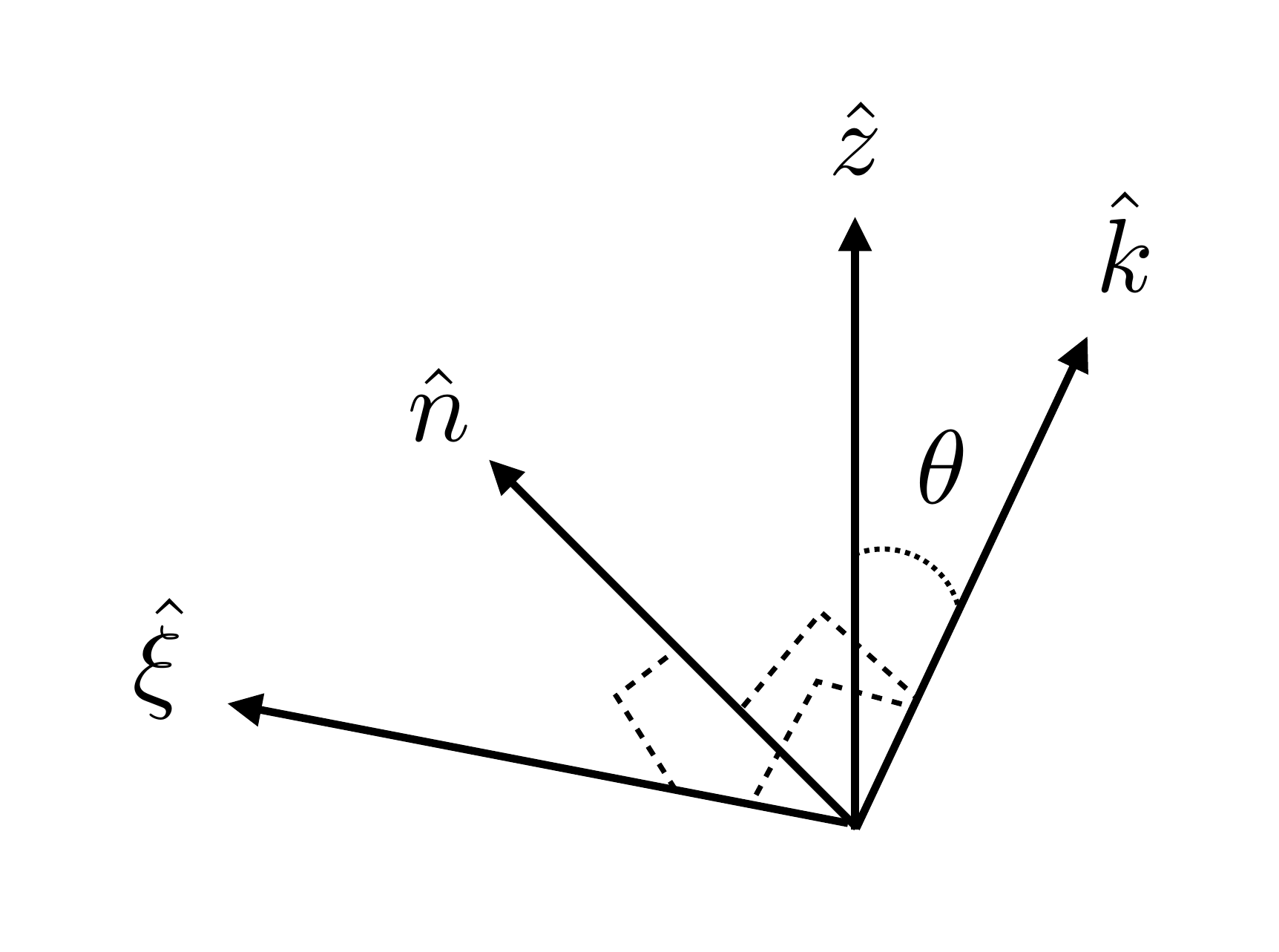}
 \caption{The triad of orthonormal vectors $\{ {\hat k}, {\hat n}, {\hat \xi} \}$ and
 the direction of the electric field along ${\hat z}$. The vectors ${\hat k}$, ${\hat \xi}$
 and ${\hat z}$ are in a plane while ${\hat n}$ is normal to the plane. }
\label{vectors}
\end{figure}

The variables $\bfp^a$ have a natural decomposition in a basis of spatial vectors
$\{ {\hat k}, {\hat n}, {\hat \xi} \}$ (see Fig.~\ref{vectors}) where 
\be
\hatk = \frac{\bfk}{k}, \ \ \hatn = \hatxi \times \hatk, \ \ 
\hatxi = \frac{\hatz - c \hatk}{s}
\ee
For convenience, we have denoted $c\equiv \hatk \cdot \hatz =\cos\theta$ and 
$s = |\hatz \times \hatk |=\sin\theta$ where $\theta$ is the angle between $\hatk$ and $\hatz$. 
Then we have the useful relation
\be
\hatz = s \hatxi + c \hatk .
\ee

Next write
\be
{\bf p}^a = \alpha_a \hatk + \beta_a \hatn + \gamma_a \hatxi
\ee
The $\{\alpha_a\}$ modes,
with polarization in the $\hatk$ direction, are longitudinal. The $\{\beta_a\}$ modes are polarized in the 
$\hatn$ direction and are transverse and also polarized orthogonal to the electric field. The $\{\gamma_a\}$
modes are transverse and polarized in the $\hatxi$ direction which is
at an angle $\pi/2 -\theta$ to the electric field. We shall call $\{\alpha_a\}$ the
``longitudinal'' (L) modes, the $\{\beta_a\}$ as the ``transverse-orthogonal'' (TO) modes, and the
$\{\gamma_a\}$ as the ``transverse-nonorthogonal'' (TN) modes. Note that for $\theta=0$, the
TN modes are also orthogonal to the electric field and coincide with TO modes.

The constraints and equations of motion in 
\eqref{consp1}-\eqref{consp3} and 
\eqref{eomp1}-\eqref{eomp3} can be written
in terms of the 9 functions $\{\alpha_a,\beta_a,\gamma_a \}$. The constraints are,
\ba
&&\omega k \alpha_1-i\Omega k \alpha_2 + \epsilon\Omega c \alpha_3 +\epsilon\Omega s \gamma_3=0 
\label{constraints1}
\\
&&\omega k \alpha_2+i\Omega k \alpha_1-i\epsilon \omega c \alpha_3 -i\epsilon \omega s \gamma_3=0 
\label{constraints2}
\\
&& \hskip -1 cm
\omega k \alpha_3+i\epsilon\omega c \alpha_2+i\epsilon\omega s\gamma_2-2\epsilon\Omega c \alpha_1
-2\epsilon\Omega s \gamma_1=0
\label{constraints3}
\ea
Note that the constraints do not involve the $\{\beta_a\}$ functions.

The equations of motion are
\ba
&& (-\omega^2-\Omega^2)\alpha_1 + i2\omega\Omega\alpha_2=0 
\label{alphaeqs1} \\
&& \hskip -1.25 cm
(-\omega^2-\Omega^2+\epsilon^2 s^2)\alpha_2 -i2\omega\Omega\alpha_1-\epsilon^2 c s\gamma_2 \nn \\
&& \hskip 4 cm
 +i\epsilon k s \gamma_3 =0 
 \label{alphaeqs2}\\
&& (-\omega^2+\epsilon^2 s^2) \alpha_3 - i\epsilon k s\gamma_2 -\epsilon^2 c s \gamma_3 =0
\label{alphaeqs3}
\ea
\ba
&& (-\omega^2+k^2-\Omega^2)\beta_1 + i 2\omega\Omega \beta_2 =0 
\label{betaeqs1} \\
&& \hskip -0.75 cm
 (-\omega^2+k^2-\Omega^2+\epsilon^2)\beta_2 -i2\omega\Omega \beta_1 -i2\epsilon k c \beta_3=0
 \label{betaeqs2} \\
&& (-\omega^2 +k^2+\epsilon^2)\beta_3 + i2\epsilon k c \beta_2 = 0
\label{betaeqs3}
\ea
\ba
&& (-\omega^2+k^2-\Omega^2)\gamma_1+i2\omega\Omega \gamma_2=0 
\label{gammaeqs1} \\
&& \hskip -0.75 cm
(-\omega^2+k^2-\Omega^2+\epsilon^2 c^2) \gamma_2 - i2\omega\Omega \gamma_1 +
i\epsilon k s \alpha_3 \nn \\
&& \hskip 3 cm - \epsilon^2 c s \alpha_2 - i 2\epsilon k c \gamma_3 =0 
\label{gammaeqs2} \\
&& \hskip -1 cm 
(-\omega^2+k^2+\epsilon^2 c^2) \gamma_3 - i \epsilon k s \alpha_2 + i2\epsilon k c \gamma_2 \nn \\
&& \hskip 4 cm
-\epsilon^2 s c \alpha_3 = 0
\label{gammaeqs3}
\ea

The $\{\beta_a\}$ functions do not appear in the constraint equations, nor do they
depend on the $\{\alpha_a,\gamma_a\}$. Hence they can be treated separately.
In the next subsection we will first consider the $\{\beta_a\}$ problem and in the
following subsection come to the more complicated $\{\alpha_a,\gamma_a\}$ problem.

\section{TO modes ($\beta_a$)}
\label{beta}

The equations for $\beta_a$ can be written as a matrix equation: $MX=0$,
\be
M=
\begin{pmatrix}
-\kappa^2-\Omega^2 & i2\omega\Omega & 0 \\
-i2\omega\Omega & -\kappa^2-\Omega^2+\epsilon^2 & -i2 \epsilon c k \\
0 & +i2 \epsilon c k & -\kappa^2+\epsilon^2
\end{pmatrix} 
\label{MTO}
\ee
where $\kappa^2 \equiv \omega^2-k^2$ and $X^T=(\beta_1,\beta_2,\beta_3)$.

To find the eigenvalues, we set the determinant of the matrix to zero. This yields a cubic equation 
in $\lambda \equiv \omega^2$,
\ba
&& \hskip -0.9 cm
P(\lambda ) \equiv
 \lambda^3 - \lambda^2 (3k^2+2\epsilon^2+2\Omega^2) \nn \\
&& \hskip -0. cm
+\lambda (3k^4+4\epsilon^2 s^2 k^2 + \epsilon^4 + \epsilon^2 \Omega^2 + \Omega^4 ) \nn \\
&& \hskip -0.9 cm
- (k^2-\Omega^2) [ (k^2+\epsilon^2 -\Omega^2)(k^2+\epsilon^2) - 4 \epsilon^2 c^2 k^2 ] = 0
\label{lambdacubic}
\ea
The cubic equation can be solved explicitly to obtain the eigenvalues, however the expressions
are opaque. We get more insight by considering a different approach.

The cubic equation in \eqref{lambdacubic} will have three roots and can be written as
\be
(\lambda-\lambda_1)(\lambda-\lambda_2)(\lambda-\lambda_3) = 0
\ee
Note that the roots $\lambda_1$, $\lambda_2$ and $\lambda_3$ for $\bfk$ and $-\bfk$ are 
identical since $k^2$, $c^2$ and $s^2$ are unchanged due to the sign flip. Hence, for example,
$\lambda_{1,\bfk} = \lambda_{1,-\bfk}$.
Together with the reality condition of \eqref{reality} this relation implies,
\be
\omega_{1,\bfk}^2 = \omega_{1,-\bfk}^2 = (\omega_{1,\bfk}^*)^2.
\ee
Hence eigenfrequencies of physical modes satisfy 
\be
\omega_\bfk = \pm \omega_\bfk^*.
\ee
{\it i.e.} physical eigenfrequencies are purely real or purely imaginary. In terms of
$\lambda$, only the real roots of \eqref{lambdacubic} are of physical interest.

\begin{figure}
\includegraphics[width=0.4\textwidth,angle=0]{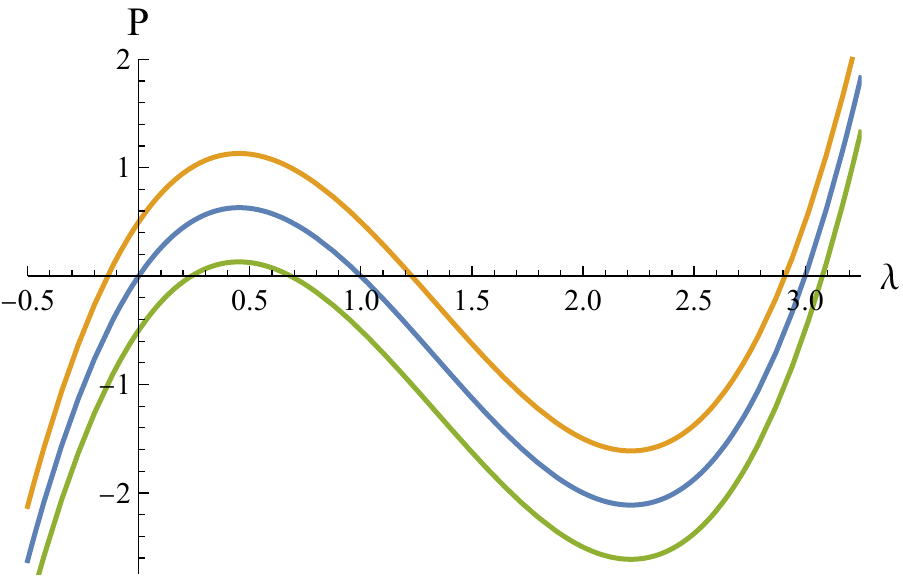}
 \caption{The cubic curve ${\tilde P}(\lambda)$ (the middle curve) and the cubics
 $P(\lambda)$ for $C>0$ (upper curve) and for $C<0$ (lower curve). The zero 
 root of ${\tilde P}$ shifts to negative $\lambda$ for $C>0$ and to positive $\lambda$
 for $C<0$. The root can become complex for sufficiently large and negative $C$.}
\label{tildeP}
\end{figure}

Next consider the polynomial as in \eqref{lambdacubic} but without the $\lambda$ independent term,
\ba
&&
{\tilde P}(\lambda) \equiv \lambda^3 - \lambda^2 (3k^2+2\epsilon^2+2\Omega^2) \nn \\
&& \hskip 2 cm
+\lambda (3k^4+4\epsilon^2 s^2 k^2 + \epsilon^4 + \epsilon^2 \Omega^2 + \Omega^4 ) \nn
\ea
Then,
\be
{\tilde P}(\lambda) \equiv \lambda (\lambda-\lambda_+) (\lambda-\lambda_-)
\ee
where $\lambda_\pm$ are obtained by solving a quadratic that involves $\bfk$ (as $k^2$
and $s^2$) and the parameters $\epsilon$ and $\Omega$. 
%
%Note that the set of roots for 
%$\bfk$ and $-\bfk$ are identical since $k^2$ and $s^2$ are unchanged due to the sign flip. 
%Together with the reality condition of \eqref{reality} we get,
%\be
%\lambda_+ = \omega_{+,\bfk}^2 = \omega_{+,-\bfk}^2 = (\omega_{+,\bfk}^*)^2.
%\ee
%Hence eigenfrequencies of physical modes satisfy 
%\be
%\omega_{+,\bfk} = \pm \omega_{+,\bfk}^*.
%\ee
%{\it i.e.} physical eigenfrequencies are purely real or purely imaginary. In terms of
%$\lambda_\pm \equiv \omega_\pm^2$, only the real roots are of physical interest.
%
We can check that the real parts of all three roots  of ${\tilde P}$
are non-negative. Therefore ${\tilde P}$ has the shape shown in Fig.~\ref{tildeP}.

\begin{figure}[h]
\includegraphics[width=0.33\textwidth,angle=0]{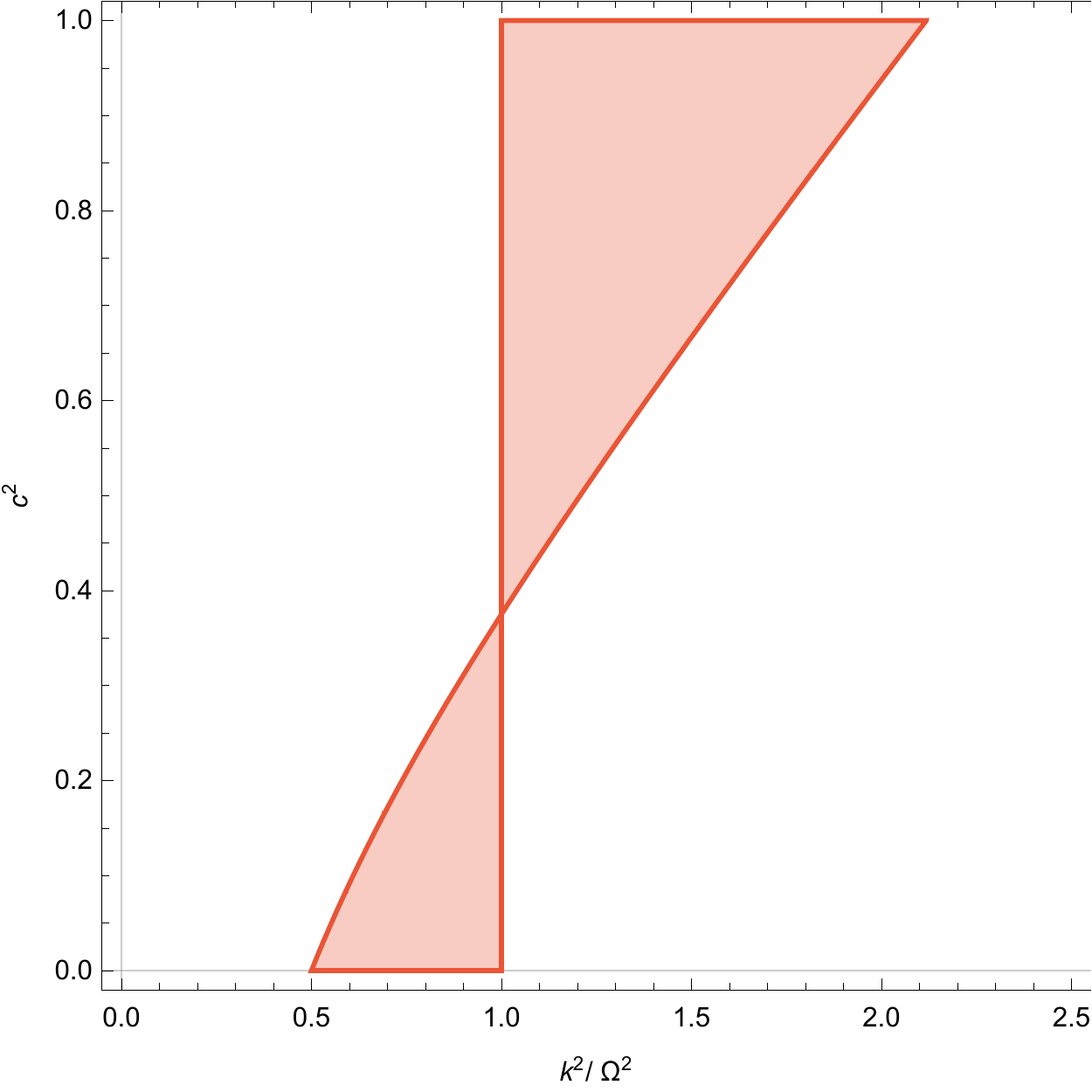}
\includegraphics[width=0.33\textwidth,angle=0]{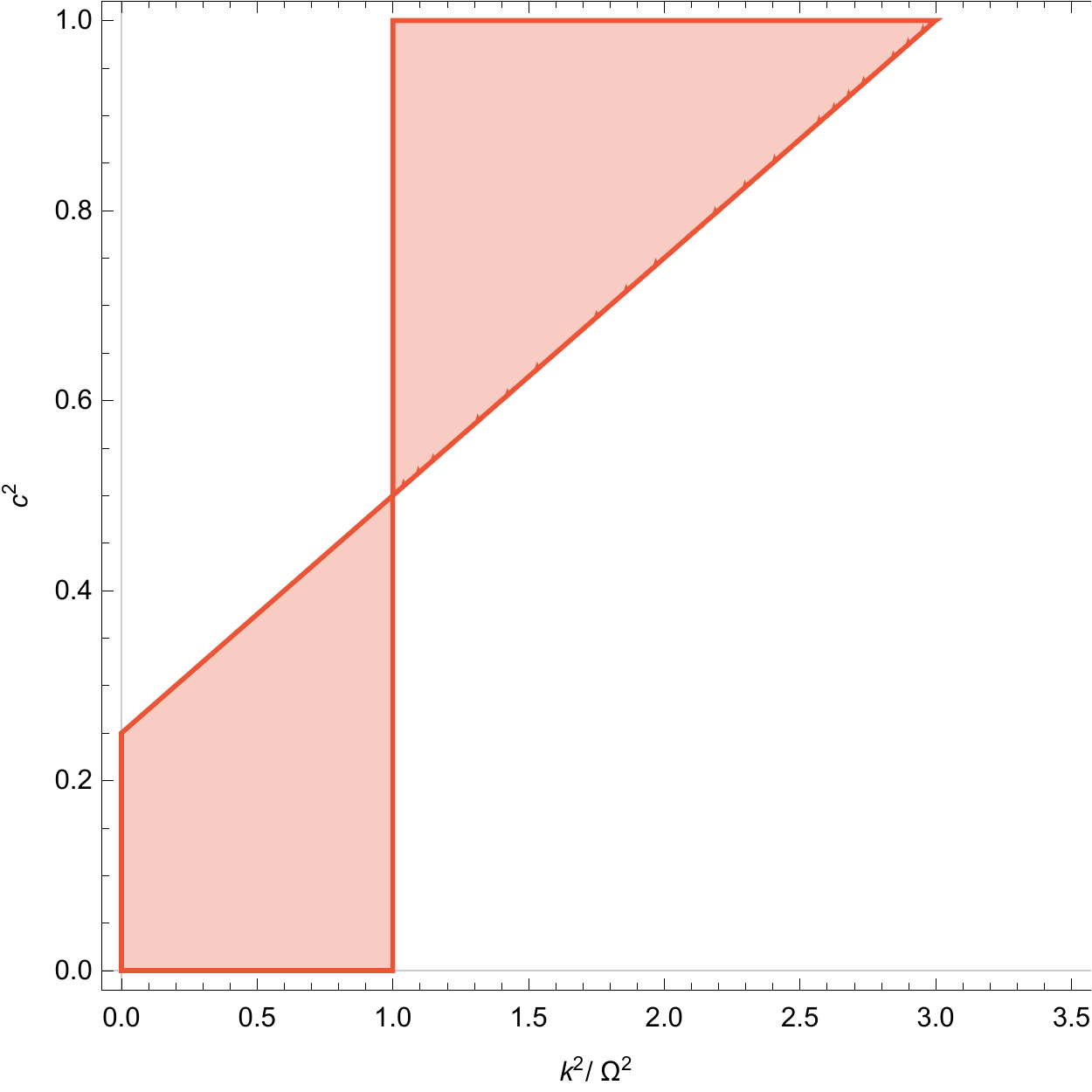}
\includegraphics[width=0.33\textwidth,angle=0]{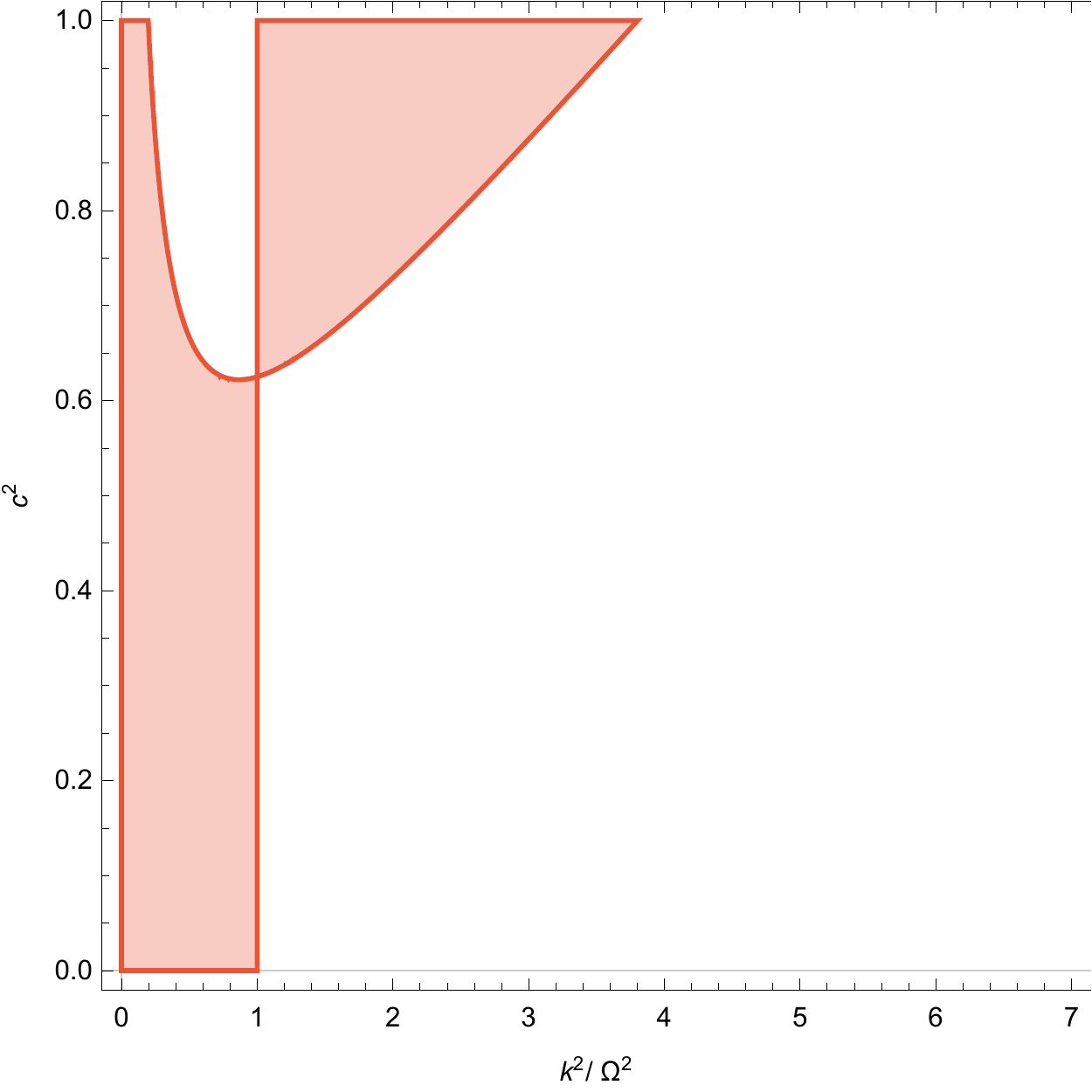}
 \caption{Stability plots for the TO modes for
 $\epsilon^2=\Omega^2/2$ (top), $\epsilon^2=\Omega^2$ (middle) and 
 $\epsilon^2=3\Omega^2/2$ (bottom). As $\epsilon$ is decreased the
 two triangular regions in the $\epsilon^2 < \Omega^2$ plot (top) shrink
 and approach the $k^2=\Omega^2$ vertical line as $\epsilon^2 \to 0$.
 }
\label{TOplot}
\end{figure}

Next let us return to the cubic in \eqref{lambdacubic} which can be written as,
\be
P(\lambda) = {\tilde P} (\lambda) + C
\ee
where
\ba
C &=& - (k^2-\Omega^2) [ (k^2+\epsilon^2 -\Omega^2)(k^2+\epsilon^2) - 4 \epsilon^2 c^2 k^2 ] \nn \\
&=& - (\kappa-\Omega^2) (\kappa-\kappa_+) (\kappa-\kappa_-)
\label{Ccubic}
\ea
where $\kappa \equiv k^2$,
\ba
\kappa_\pm &=& \frac{1}{2} \biggl [ \Omega^2 +2\epsilon^2 c_2 \nn \\
&& \hskip 0.5 cm
\pm  \sqrt{ \left ( \Omega^2 +2\epsilon^2 c_2 \right )^2 - 4\epsilon^2 (\epsilon^2-\Omega^2)}\ \biggr ]
\label{kappapm}
\ea
and $c_2 \equiv \cos(2\theta)$.
If $C > 0$, the ${\tilde P}$ curve shifts upwards (see Fig.~\ref{tildeP}) and the $\lambda=0$
root of ${\tilde P}$ shifts to the left, {\it i.e.} $\lambda < 0$. This indicates an instability. On
the other hand, if $C < 0$, there is no instability.

It is simplest to analyze the case with $\epsilon^2=\Omega^2$ for then 
\be
\kappa_\pm = \frac{\Omega^2}{2} \left [ (4c^2-1) \pm | 4c^2-1| \right ]
\ee
For $c^2 > 1/4$, $\kappa_-=0$ and $\kappa_+ = \Omega^2 (4 c^2-1)$. Then for $c^2 >1/2$
we obtain an instability ($C >0$) for $\Omega^2 < k^2 < \Omega^2 (4 c^2 -1)$
and for $1/4 < c^2 < 1/2$ modes are unstable if $\Omega^2 (4 c^2 -1) < k^2 < \Omega^2$.
For the opposite case of $c^2 < 1/4$, $\kappa_- = 4c^2-1 < 0$ and $\kappa_+=0$. Then
the instability occurs for $0 < k^2 < \Omega^2$. This explains the instability domains shown
in Fig.~\ref{TOplot} in the $\epsilon^2=\Omega^2$ case.

The analysis and results are similar for $\epsilon^2 < \Omega^2$ as is clear from Fig.~\ref{TOplot}.
With $c=0$, we find that $C > 0$ for $\Omega^2-\epsilon^2 < k^2 <  \Omega^2$.
Essentially, as $\epsilon^2$ is reduced the domain of instability shrinks towards a vertical line along
$k^2=\Omega^2$.  
An interesting limit is $\epsilon \to 0$ and $\Omega \to \infty$ with 
$\epsilon \Omega = E$ held constant. In this limit, there are no unstable TO modes.

The analysis is a bit more involved for $\epsilon^2 > \Omega^2$. First consider $c^2=0$ ($c_2=-1$).
Then \eqref{kappapm} gives $\kappa_\pm = -\epsilon^2, \, -\epsilon^2+\Omega^2$ and
both roots are for negative $\kappa$. From \eqref{Ccubic} we then see that $C > 0$ for
$0 < k^2 < \Omega^2$ as is also seen in Fig.~\ref{TOplot}. Next consider $c^2=1$ ($c_2=+1$).
Then \eqref{kappapm} gives
\be
\kappa_\pm = \frac{1}{2} \biggl [ \Omega^2 +2\epsilon^2 
\pm  \sqrt{  \Omega^4 +8 \epsilon^2 \Omega^2 }\ \biggr ] .
\ee
Now $\kappa_+ > \Omega^2$ but $\kappa_-$ may be larger or smaller than $\Omega^2$.
If $\kappa_- < \Omega^2$, then from \eqref{Ccubic} we see that 
$C > 0$ for $0<k^2 < \kappa_-$  and also for $\Omega^2 < k^2 < \kappa_+$ but
$C < 0$ for $\kappa_- < k^2 < \Omega^2$. Hence we see the gap in the instability
domain in Fig.~\ref{TOplot} at $c^2=1$. If, on the other hand, $\kappa_- > \Omega^2$,
then $C > 0$ for $0 < k^2 < \Omega^2$ and then again for $\kappa_- < k^2 < \kappa_+$.

The shapes of the unstable regions can be understood in terms of the roots of $C$ in 
\eqref{Ccubic}, namely 1 and $\kappa_\pm$. For example, in the case $\epsilon^2 > \Omega^2$,
for $c^2=0$, there is one real root ($\kappa = \Omega^2$) and 
$\kappa_\pm$ are negative.
As $c^2$ increases, $\kappa_\pm$ become complex.
At some critical value of $c^2$ the imaginary part of $\kappa_\pm$ vanishes. This is at the minimum
of the parabolic shape in the plot of Fig.~\ref{TOplot} and can occur for $k^2 < \Omega^2$ or 
$k^2 >\Omega^2$ depending
on the value of $\epsilon^2/\Omega^2$. The left edge of the parabola is given by $\kappa_-$, and the 
right edge is given by $\kappa_+$. For yet larger $c^2$, $\kappa_+$ becomes larger than $\Omega^2$.
The unstable region is when two of the factors in \eqref{Ccubic} are positive and one is negative.

\section{TN and L modes ($\{\alpha_a,\gamma_a\}$)}
\label{alphagamma}

Eqs.~\ref{alphaeqs1}-\ref{alphaeqs3} and \ref{gammaeqs1}-\ref{gammaeqs3} can be written in 
matrix form $MX=0$ with 
$X^T=(\alpha_1,\alpha_2,\alpha_3,\gamma_1,\gamma_2,\gamma_3)$ and
\begin{widetext}
\be
M =
\begin{pmatrix}
-\omega^2-\Omega^2 & i2\omega\Omega & 0 &0 &0 &0 \\
 -i2\omega\Omega & -\omega^2-\Omega^2 + \epsilon^2 s^2 & 0&0& -\epsilon^2 cs & i\epsilon sk \\
 0&0&-\omega^2+\epsilon^2 s^2 &0&-i\epsilon s k & - \epsilon^2 cs \\
 0&0&0&-\omega^2+k^2-\Omega^2 &i2\omega\Omega&0\\
 0&-\epsilon^2 cs&i\epsilon sk&-i2\omega\Omega&-\omega^2+k^2-\Omega^2+\epsilon^2 c^2&-i2\epsilon ck \\
 0&-i\epsilon sk & -\epsilon^2 cs & 0 & i2\epsilon ck & -\omega^2+k^2+\epsilon^2 c^2 
\end{pmatrix}
\label{MTNL}
\ee
\end{widetext}
The eigenvectors are also required to satisfy the constraints in \eqref{constraints1}-\eqref{constraints3}
as we will discuss further after considering some special cases.

\subsection{Special case: $c=1$}
\label{c=1}

For $c=1$, the matrix in \eqref{MTNL} becomes block diagonal in 3 blocks, the first is the
$\{\alpha_1,\alpha_2\}$ $2\times 2$
block with two degenerate eigenvalues $\omega^2 = \Omega^2$, the second block is 
the $\{\alpha_3\}$ $1\times 1$
block with eigenvalue $\omega^2=0$, and the third $\{\gamma_1,\gamma_2,\gamma_3\}$
block is given by the matrix in \eqref{MTO}
with $c=1$. Then the analysis in Sec.~\ref{beta}  for the TO modes applies immediately
(with $c=1$). This is expected since for $c=1$ there is no distinction between TO and TN modes.

The constraint equations with $c=1$ read,
\ba
&&\omega k \alpha_1-i\Omega k \alpha_2 + \epsilon\Omega  \alpha_3 =0 
\label{c1}\\
&&\omega k \alpha_2+i\Omega k \alpha_1-i\epsilon \omega \alpha_3 =0
\label{c2}\\
&&\omega k \alpha_3+i\epsilon\omega \alpha_2 -2\epsilon\Omega \alpha_1 =0
\label{c3}
\ea
Note that the $\gamma_a$ are unconstrained. On the other hand, 
\eqref{c1}-\eqref{c3} over-constrain
the eigensolutions in the $\{\alpha_1,\alpha_2\}$ sector with $\omega=\pm \Omega$,
and in the $\{\alpha_3\}$ sector with $\omega=0$, and neither of these two
eigensolutions are physically admissible.

\subsection{Special case: $c=0$}
\label{c=0}

With $c=0$, we have $s=1$, and the matrix in \eqref{MTNL} becomes block diagonal in
the $\{\alpha_1,\alpha_2,\gamma_3\}$ and the $\{\alpha_3,\gamma_1,\gamma_2\}$ blocks.
The $3\times 3$ matrix for the first block is
\be
M_1=
\begin{pmatrix}
-\omega^2-\Omega^2 & i2\omega\Omega & 0 \\
-i2\omega\Omega & -\omega^2-\Omega^2+\epsilon^2 & i \epsilon k \\
0 & -i \epsilon k & -\omega^2+k^2
\end{pmatrix} 
\label{M1}
\ee
with constraint,
\ba
&&\omega k \alpha_1-i\Omega k \alpha_2  +\epsilon\Omega  \gamma_3=0 
\label{M1c1}
\\
&&\omega k \alpha_2+i\Omega k \alpha_1 -i\epsilon \omega  \gamma_3=0 
\label{M1c2}
\ea
and the matrix for the second block is
\be
\hskip -0.35 cm
M_2 =
\begin{pmatrix}
 -\omega^2+\epsilon^2 &0&-i\epsilon k \\
 0&-\omega^2+k^2-\Omega^2 &i2\omega\Omega \\
 i\epsilon k&-i2\omega\Omega&-\omega^2+k^2-\Omega^2 
\end{pmatrix}
\label{M2}
\ee
with constraint,
\be
\omega k \alpha_3+i\epsilon\omega \gamma_2-2\epsilon\Omega \gamma_1=0.
\label{M2c1}
\ee

We now discuss these $3\times 3$ blocks separately.

\subsubsection{$\{\alpha_1,\alpha_2,\gamma_3\}$ block}
\label{a1a2g3}

In this block, gauge fields of the first two colors are oscillating in the longitudinal
direction, whereas the third has amplitude in the transverse direction and
orthogonal to the background electric field.

A straight-forward procedure would be to first solve the eigenproblem for $M_1$ and
then check for the eigenvectors that satisfy the constraints. However we find it simpler
to first solve the constraints \eqref{M1c1}-\eqref{M1c2} and then deal with the eigenproblem.

The constrains in \eqref{M1c1}-\eqref{M1c2} can be used to eliminate two of the
three variables, say $\alpha_2$ and $\gamma_3$, while the third variable can be
absorbed in the normalization of the resulting eigenvector. Hence we seek an
eigenvector of the form,
\be
V_1 \equiv \begin{pmatrix} \alpha_1\\ \alpha_2 \\ \gamma_3 \end{pmatrix}
= 
\begin{pmatrix} 2\epsilon \omega \Omega \\ -i \epsilon (\omega^2+\Omega^2) \\ 
k(-\omega^2+\Omega^2) \end{pmatrix}
\ee
Insertion in $M_1 V_1=0$ shows that there is no solution for $\omega$ for $\epsilon \ne 0$,
$k\ne 0$. Hence these modes are over-constrained and absent.
For $k=0$, $M_1 V_1=0$ gives
\be
(\omega^2-\Omega^2)^2-\epsilon^2 (\omega^2+\Omega^2) =0
\label{M1V1condition}
\ee
which has the roots
\be
\omega_\pm^2 = \Omega^2+\frac{\epsilon^2}{2} \pm 
\sqrt{ \left (  \Omega^2+\frac{\epsilon^2}{2} \right )^2 + \Omega^2 (\epsilon^2 -\Omega^2) }
\label{M1V1roots}
\ee
Therefore $\omega_-^2 < 0$ if and only if $\Omega^2 < \epsilon^2$ and $k=0$.

\subsubsection{$\{\alpha_3,\gamma_1,\gamma_2\}$ block}
\label{a3g1g2}

In this block, the gauge field of the third color oscillates in the longitudinal
direction, whereas the first two colors oscillate transversely and
orthogonal to the background electric field.

Now the constraint \eqref{M2c1} reduces the eigenvector to be of the form,
\be
V_2 \equiv \begin{pmatrix} \alpha_3\\ \gamma_1 \\ \gamma_2 \end{pmatrix}
= 
\begin{pmatrix} 2\epsilon \Omega \alpha_3 \\ \omega (k \alpha_3+i\epsilon \gamma_2) \\ 
2 \epsilon \Omega \gamma_2 \end{pmatrix} .
\ee
Imposing $M_2 V_2 =0$ we find the solution
$\omega^2=\epsilon^2$ provided $k^2=\epsilon^2+\Omega^2$, and this
mode is stable. For $k=0$ there is a solution with $\omega^2=\epsilon^2$.

An unusual feature of the first of these two modes is that it has non-trivial spatial
dependence but it exists only for a fixed value of 
$k=\sqrt{\epsilon^2+\Omega^2}$ and the direction of $\bfk$ is perpendicular to the 
electric field, while the mode is polarized along the electric field and in the
longitudinal direction. This mode represents fluctuations in the homogeneity
of the electric field but with a definite wavelength.

\subsection{Special case: $k^2 \gg \Omega^2, \epsilon^2$}
\label{klarge}

We now consider the ultraviolet limit $k^2 \gg \Omega^2, \epsilon^2$. The
constraints \eqref{constraints1}-\eqref{constraints3} now give
\be
\alpha_1=0=\alpha_2=\alpha_3
\ee
and only the $\{\gamma_a\}$ represent physical modes with the dispersion
relation $\omega^2=k^2$. There are no unstable modes in this limit.

\subsection{Special case: $k \to 0$}
\label{k=0}

The problem simplifies in the $k \to 0$ limit as the $\{\alpha_3,\gamma_3\}$ block
decouples from the $\{\alpha_1,\alpha_2,\gamma_1,\gamma_2\}$ block. 

The $2\times 2$ matrix for the $\{\alpha_3,\gamma_3\}$ block is
\be
M_3 = \begin{pmatrix}
-\omega^2 +\epsilon^2 s^2 & - \epsilon^2 cs \\
- \epsilon^2 cs & -\omega^2 +\epsilon^2 c^2
\end{pmatrix}
\ee
and the constraint reduces to $c \alpha_3 + s \gamma_3=0$.
$M_3$ has eigenvalues $\omega^2=0$ and $\omega^2=\epsilon^2$ but only the
latter is consistent with the constraint.
Thus there are no unstable modes in the $\{\alpha_3,\gamma_3\}$ block.

The $4\times 4$ matrix for the $\{\alpha_1,\alpha_2,\gamma_1,\gamma_2\}$ block is
\begin{widetext}
\be
M_4 = \begin{pmatrix}
-\omega^2 -\Omega^2 & i2\omega\Omega &0 &0 \\
-i2\omega\Omega & -\omega^2 -\Omega^2 +\epsilon^2 s^2 & 0 & - \epsilon^2 c s \\
0 & 0 & -\omega^2 -\Omega^2 & i2\omega\Omega \\
0 & - \epsilon^2 c s &  -i2\omega\Omega & -\omega^2 -\Omega^2 +\epsilon^2 c^2 
\end{pmatrix}
\ee
\end{widetext}
and the constraint is 
\be
i\omega (c \alpha_2+ s\gamma_2)-2\Omega (c \alpha_1 + s \gamma_1)=0.
\label{constraintk0}
\ee
We solve \eqref{alphaeqs1} and \eqref{gammaeqs1} with $k=0$ to get
\be
\alpha_2 = -i \frac{(\omega^2+\Omega^2)}{2\omega\Omega} \alpha_1, \ \ 
\gamma_2 = -i \frac{(\omega^2+\Omega^2)}{2\omega\Omega} \gamma_1,
\label{a2a1g2g1}
\ee
which, together with \eqref{constraintk0}, gives
\be
(\omega^2-3\Omega^2)(c\alpha_1+s\gamma_1)=0.
\ee
Therefore to satisfy the constraint we must either have $\omega^2=3\Omega^2 >0$
or $c\alpha_1+s\gamma_1=0$. 

Evaluation of the determinant of $M_4$ on Mathematica gives,
\be
{\rm Det}(M_4) = (\omega^2-\Omega^2)^2 [ (\omega^2-\Omega^2)^2-\epsilon^2 (\omega^2+\Omega^2)]
\ee
This has the root $\omega^2=3\Omega^2$ but only if $\Omega^2=\epsilon^2$. In any case,
$\omega^2 = 3\Omega^2 >0$ and implies a stable mode. So we now focus on the other
case, namely 
\be
c\alpha_1+s\gamma_1=0.
\label{casg}
\ee
Combining \eqref{casg} with \eqref{a2a1g2g1}, and ignoring an overall normalization factor,
the Gauss constraint forces us to only consider the eigenvector,
\be
\hskip -0.2 cm
V_4^T = \left (2\omega\Omega s, -i {(\omega^2+\Omega^2)} s, 
                     -2\omega\Omega c, i {(\omega^2+\Omega^2)} c \right ) .
\label{V4}
\ee
Requiring $M_4 V_4 =0$ leads once again to \eqref{M1V1condition} and
to the roots in \eqref{M1V1roots}.
Therefore $\omega_-^2 < 0$ if and only if $\Omega^2 < \epsilon^2$ and $k=0$ and the unstable 
eigenmode can be found by setting $\omega=\omega_-$ in \eqref{V4}. 

\subsection{Special case: $\Omega = E/\epsilon$, $\epsilon \to 0$}
\label{limitingcase}

In Sec.~\ref{beta} we have seen that there are no unstable TO modes with $\Omega = E/\epsilon$ 
and $\epsilon \to 0$. Now we consider the TN and L modes in this regime.

With $\epsilon \to 0$, the matrix $M$ in \eqref{MTNL} takes on a simple block diagonal form. 
The $\{\alpha_1,\alpha_2 \}$ block has two degenerate eigenvalues $\omega^2 = \Omega^2$;
the $\{\alpha_3\}$ block has eigenvalue $\omega^2=0$; 
the  $\{\gamma_1,\gamma_2 \}$ block has eigenvalues $\omega^2 = (k \pm \Omega)^2$; 
and the $\{\gamma_3\}$ block has eigenvalue $\omega^2=k^2$. The corresponding
eigenvectors can be inserted into equations \eqref{constraints1}-\eqref{constraints3} to 
check if the Gauss constraints are satisfied. However, since none of the eigenvalues for 
$\omega^2$ are negative, it is clear that there are no unstable TN and L modes for these
limiting parameters.

\section{Conclusions}
\label{conclusions}

We have considered the stability of a homogeneous electric field background in pure 
SU(2) gauge theory. The gauge fields underlying the electric field are taken to be of the form in
\eqref{Amu} and not of the Maxwell type: $A_i^a = -E t \delta^{a3}\delta_{iz}$. This 
is because gauge fields of the Maxwell
type are unstable to Schwinger pair production while the gauge fields in \eqref{Amu}
are protected from decay due to this process~\cite{Vachaspati:2022ktr}. However,
the gauge fields in \eqref{Amu} are not solutions of the vacuum classical equations 
of motion; instead non-vanishing currents are necessary. There are two ways to
explain these non-vanishing currents. The first is that they are due to classical external
sources in which case they are simply postulated. The second way is that the
classical equations of motion should be replaced by equations that take quantum
effects into account and these ``effective classical equations'' can contain sources.
For example, in the semiclassical approximation quantum fluctuations provide 
current sources for the background~\cite{Vachaspati:2022ktr},
\ba
j^{\mu a} &\equiv& \epsilon^{abc} \langle \partial_\nu q^{\nu b} q^{\mu c}  
- q^{\nu b} \partial^\mu q_\nu^c + 2 q^{\nu b} \partial_\nu q^{\mu c} \rangle_R \nn \\
&& 
+ A_\nu^b \langle q^{\nu a}q^{\mu b} - 2 q^{\nu b} q^{\mu a} \rangle_R
+ A_\nu^a \langle q^{\nu b} q^{\mu b} \rangle_R \nn \\
&&
+ A^{\mu b} \langle q^b_\nu q^{\nu a} \rangle_R - A^{\mu a} \langle q_\nu^b q^{\nu b} \rangle_R
\label{jmua}
\ea
where $q^a_\mu$ are the quantum fluctuations in the background $A^a_\mu$ and
$\langle \cdot \rangle_R$ denotes a renormalized expectation value taken in the
quantum state of $q^a_\mu$. For stable modes, the quantum state might be given
by simple harmonic oscillator states for each of the eigenmodes of $q^a_\mu$.
However the quantum state of unstable modes will not be simple
harmonic oscillator states which is why it is important to perform a stability
analysis.
We will comment further on the unstable modes after summarizing our results.

The gauge field background in \eqref{Amu} is described by two parameters: $\epsilon$ 
and $\Omega$. 
The electric field strength is given by $E = \epsilon\Omega$. The results of the fluctuation 
analysis depend on whether $\epsilon^2 > \Omega^2$ or $\epsilon^2 \le \Omega^2$.
The fluctuations naturally split into ``TO modes"" that are transverse to the wavevector $\bfk$
and orthogonal to the background electric field, ``TN modes'' that are transverse
to $\bfk$ but not orthogonal to the electric field, and ``L modes'' that are in the
longitudinal direction. 

The TO modes decouple from the TN and L modes.
The stability analysis of Sec.~\ref{beta} shows that TO modes are stable except
in a range of $k^2$ that depends on the angle $\theta$ between the electric field and the
wavevector $\bfk$. The instability regions depend on the background parameters and
are plotted in Fig.~\ref{TOplot}. There are two important results emerging from our
analysis. The first is that the region of parameter space $(k^2, c^2)$ ($c=\cos\theta$)
where unstable modes exist depends on the relation between $\epsilon^2$ and $\Omega^2$. 
The instability region is smaller when $\epsilon^2 < \Omega^2$ and shrinks to zero as
$\epsilon^2 \to 0$. Note that the electric field strength is given by $E=\epsilon\Omega$
and can be held fixed in the limit by also taking $\Omega \to \infty$. The second is
that unstable modes exist only for small values of $k^2$. For example, for $\epsilon=\Omega$,
there are no unstable modes for $k^2 > 3\epsilon^2$ for any value of $c^2$.

The TN and L modes are coupled in general and the analysis is more involved
than for the TO modes. In Sec.~\ref{alphagamma} we discuss the stability of
these modes in various parameter regimes. The special cases of $\theta=0$
and $\theta=\pi/2$ are considered. For $\theta=0$ the analysis is identical
to that of TO modes, while for $\theta=\pi/2$ 
there is an instability if $\epsilon^2 > \Omega^2$ and $k=0$.
There is also a special stable mode that corresponds to oscillations of 
the background electric field orthogonal to its direction, similar to a sound wave.
We have also considered the special case of large $k^2$ and here the
modes are simply those of free massless waves with dispersion $\omega^2=k^2$.
Finally, we examine the
$\epsilon \to 0$ limit with $E=\epsilon\Omega$ held fixed and show that
there are no unstable TN and L modes, just as there are no unstable TO modes
in this limit.

As mentioned in Sec.~\ref{introduction}, we were motivated to perform this stability
analysis because confining strings in QCD are expected to be stable.
The electric fields we have considered
as backgrounds do not excite Schwinger pair production but, as we have
seen, have classical instabilities for certain infrared modes. How do these 
classical instabilities
impact the possibility that the electric fields we have considered are responsible
for confining strings? The first point we note is that there are no 
instabilities in the limit of $\epsilon \to 0$ and $E = \epsilon\Omega$ fixed.
So it could be that the electric field in a confining string corresponds to this set 
of parameters. Then there are no unstable modes and the quantum state for
each mode is that of a simple harmonic oscillator. The second point is that the 
instabilities we have found are
for a {\it homogeneous} electric field and only occur for small values of $k^2$,
($k^2 \lesssim \epsilon^2$ for $\epsilon^2 > \Omega^2$)
that is, on large length scales. In contrast, the electric flux in a string only has 
support in a finite area -- the string cross-section -- and we do not expect
any unstable modes on length scales larger than the thickness
of the string. (Though there is still the 
question of the infinite extent of the string along the electric field direction and
whether the instabilities for $\theta=0$ will survive.) It would be worthwhile
performing an explicit stability analysis for a flux tube configuration such 
as~\cite{Vachaspati:2022ktr},
\be
A^\pm_\mu = -\epsilon  e^{\pm i \Omega t} \, f(r) \, \partial_\mu z , \ \ 
A^3_\mu =0,
\label{bkgndpmf}
\ee
where $f(r)$ is a profile function for the string and $r \equiv \sqrt{x^2+y^2}$ is the 
cylindrical radial coordinate.
Another interesting question is if the homogeneous electric field background we 
have considered is unstable towards forming an Abrikosov lattice~\cite{Abrikosov:1956sx} 
of electric flux tubes. After all we have identified certain unstable modes with spatial
dependence that is orthogonal to the background homogeneous electric field.

\acknowledgements
We are grateful to Jan Ambjorn, Guy Moore, Stanislaw Mr\'owczy\'nski, Igor Shovkovy and 
Raju Venugopalan for comments. T. V. thanks the CCPP at NYU for hospitality where some 
of this work was done.
This work was supported by the U.S. Department of Energy, Office of High Energy 
Physics, under Award No.~DE-SC0019470.

%\appendix
%
%\section{Reality condition}
%\label{realityapp}
%
%We first define,
%\be
%q_\mu^\pm \equiv q_\mu^1 \pm i q_\mu^2
%\ee
%Since $q_\mu^a$ are real, this implies $q_\mu^+ = (q_\mu^-)^*$.
%
%Next consider the spatial Fourier transform,
%\be
%q_\mu^a (t,\bfx) = \int \frac{d^3k}{(2\pi)^3} e^{i\bfk \cdot \bfx} s_\mu^a (t,\bfk)
%\ee
%and
%\be
%q_\mu^\pm (t,\bfx ) = \int \frac{d^3k}{(2\pi)^3} e^{i\bfk \cdot \bfx} s_\mu^\pm (t,\bfk)
%\ee
%with $ s_\mu^\pm (t,\bfk)= s_\mu^1 (t,\bfk) \pm i s_\mu^2 (t,\bfk)$.
%

\newpage

\bibstyle{aps}
\bibliography{paper}

\end{document}